\documentclass[pra,showpacs,showkeys]{revtex4} \usepackage{latexsym}
\usepackage{amsfonts} \usepackage{amsmath,revsymb} 

\newcommand{\kt}[1]{\ensuremath{|#1\rangle}}
\newcommand{\br}[1]{\ensuremath {\langle #1|}}
\newcommand{\bk}[2]{\ensuremath {\langle #1|#2 \rangle}}

\newcommand{\vp}{{\bf p}}

\newcommand{\su}{{\rm SU}(2)}

\begin{document}

\title{Dynamical entanglement in particle scattering}

\author{N.L.~Harshman}
\affiliation{Department of Computer Science, Audio Technology and Physics\\
4400 Massachusetts Ave., NW \\ American University\\ Washington, DC 20016-8058}

\begin{abstract}

This paper explores the connections between particle scattering and quantum information theory in the context of the non-relativistic, elastic scattering of two spin-1/2 particles.  An untangled, pure, two-particle in-state is evolved by an S-matrix that respects certain symmetries and the entanglement of the pure out-state is measured.  The analysis is phrased in terms of unitary, irreducible representations (UIRs) of the symmetry group in question, either the rotation group for the spin degrees of freedom or the Galilean group for non-relativistic particles.  Entanglement may occurs when multiple UIRs appear in the direct sum decomposition of the direct product in-state, but it also depends of the scattering phase shifts.
\end{abstract}

\maketitle

\section{Introduction}

Entanglement has become to be seen as a resource for certain quantum information processes like computation~\cite{shor}, encryption~\cite{BB84}, and teleportation~\cite{teleport}.  Quantifying this entanglement is a necessary step to manipulating this resource.  There are many results for entanglement measures of two-party, finite dimensional systems (see Ref.~\cite{plenio05} for a review),  but much remains unknown about entanglement measures of systems that are multi-party, are infinite-dimensional, and/or have continuous degrees of freedom.  In this paper we consider entanglement in the spin degrees of freedom of two spin-1/2 scattering particles.  This problem will allow us to use the mathematical results for two-qubit systems, such as the entropy of entanglement, in a system that also has other degrees of freedom.

This paper addresses three issues that arise when considering entanglement in scattering systems.  How do the boundary conditions of a scattering experiment affect the amount of entanglement?  How do the symmetries and invariances of the interaction restrict the maximum entanglement for given in-states?  What role do the degrees of freedom like momentum, whose entanglement is not considered, play in determining the final entanglement in the spin degrees of freedom?

The scattering system will be isolated from external fields, and the individual particles in the in-state will also be assumed to be free.  As a result, Galilean symmetry will assumed for the individual particles in the in- and out-states, and also for the whole, isolated system.  In particular, the S-matrix, the scalar scattering operator, will be diagonal in the generators of the Galilean group of the total system.  Because of these symmetries (and possibly other symmetries of the interaction), it will be useful to work with unitary irreducible representations (UIRs) of the Galilean group and its subgroup, the rotations.  Spin systems correspond to UIRs of $\su$ and non-relativistic particles correspond to UIRs of the Galilean group (extended by the central charge of mass)~\cite{gal,nraqft}.

The product of the UIRs associated with each particle is no longer irreducible, but can be decomposed into direct sum of UIRs (Clebsch-Gordan series) that may have a certain multiplicity.  For example, when the direct product of the spin-1/2 UIRs of $\su$ are decomposed into a direct sum, UIRs with $s=0$ and $s=1$ appear once.  For the Galilean (and Poincar\'e) group, this reduction process is often called partial wave analysis.  To summarize the main results, it appears that dynamic entanglement can only occur if the in-state has components in multiple UIRs of the direct sum decomposition.

The first part of this paper will discuss the ``scattering'' of systems with only spin degrees of freedom.  Dynamical entanglement in this context will be constrained by overall rotational symmetry of the interaction.  Then the case of dynamical entanglement of spin degrees of freedom in non-relativistic scattering of spin-1/2 particles will be treated.  For the case of central forces between the particles, we find that in each partial wave of orbital angular momentum, dynamical entanglement of the spin degrees of freedom works the same way as the simpler case of spin only.

\section{Dynamical Entanglement of Spin Degrees of Freedom}

First we consider pure states of two spin $s=1/2$ systems.  The pure state of an individual particle is as an element of the two-dimensional representation space of the rotation group $\su$.  
In the basis of this representation space, $\kt{s \chi}$ is a basis vector where $s$ denotes total spin angular momentum and $\chi=\pm$ the (sign of the) component of angular momentum in some direction.  In this basis, the representation of a transformation $u\in \su$ is denoted
\begin{equation}\label{trans}
U(u)\kt{s\chi}=\sum_{\chi'=-s}^s D^{s}_{\chi'\chi}(u) \kt{s\chi'}
\end{equation}

Using the direct product basis kets $\kt{\chi_1 \chi_2}=\kt{s_1\chi_1}\otimes \kt{s_2\chi_2}$, a general state can be written
\begin{equation}
\psi = \sum_{\chi_1\chi_2} c_{\chi_1\chi_2}\kt{\chi_1 \chi_2}.
\end{equation}
The direct product representation can be decomposed into UIRs using the Clebsch-Gordan series.  For $\su$, each UIR in the direct sum appears only once,
\begin{equation}
D^{s_1} \otimes D^{s_2}  = \sum_{s = |s_1 - s_2|}^{s_1 + s_2} D^s.
\end{equation}
The direct product and direct sum basis vectors are connected by the Clebsch-Gordan coefficients (CGCs) for $\su$:
\begin{equation}
\kt{\chi_1 \chi_2} = \sum_{s\chi} \bk{s\chi}{\chi_1 \chi_2} \kt{s\chi}.
\end{equation}
By convention the CGCs $\bk{s\chi}{\chi_1 \chi_2}$ of $\su$ are chosen to be real and their properties are well-known.

Since we will be considering pure states of bipartite systems with finite degrees of freedom, the entropy of entanglement~\cite{entent}
\begin{equation}\label{ent}
E(\kt{\psi}) = S(\rho_1) = S(\rho_2)
\end{equation}
 is an accepted measure,
where $\rho_1 = \mathrm{tr}_2 [\kt{\psi}\br{\psi}]$ is the density matrix for particle 1 that remains after a partial trace over particle 2, and  $S(\rho)= - \mathrm{tr}[\rho\log_2\rho]$ is the Von Neumann entropy of the density matrix $\rho$ (taken base two for normalization).  Various other definitions for the entanglement of mixed states (distillation, formation, etc.) all reduce to this in the case of pure states.\cite{plenio05}

This entanglement measure takes its minimal value on any pure, normalized, product state of the form
\begin{equation}\label{prod}
\kt{\Phi_{12}} = \kt{\phi_1}\otimes\kt{\phi_2} = \sum_{\chi_1\chi_2}a_{\chi_1}b_{\chi_2}\kt{\chi_1 \chi_2}
\end{equation}
For such states, the reduced density matrix is $\rho_1 = \kt{\phi_1}\br{\phi_1}$ and the entanglement is $E(\kt{\Phi_{12}})=0$.  These states are separable.  

The boundary conditions of scattering experiments require that the in-states be separable.
Since every pure state of a spin-1/2 system can be described by a polarization vector $\hat{n}_i$, without loss of generality, we will choose the $z$-direction to be the polarization direction $\hat{n}_1$ and the $x$-direction to lie in the plane containing $\hat{n}_1$ and $\hat{n}_2$ such that $\hat{x}\cdot\hat{n}_2 = \theta \in [0,\pi)$.  Choosing this coordinate system, any pure, product state can be written as
\begin{equation}\label{in-ss}
\Phi = \cos\theta\kt{++} + \sin\theta\kt{+-}
\end{equation}
and in the direct sum basis
\begin{equation}\label{in-jm}
\Phi = \cos\theta\kt{11} + \frac{1}{\sqrt{2}}\sin\theta\left( \kt{00} + \kt{10}\right).
\end{equation}

The maximum entanglement value $E= 1$ occurs for any of the ``magic basis'' kets
\begin{eqnarray}
\kt{\Phi_{EPR\pm}} &=& \frac{1}{\sqrt{2}}\left(\kt{+-}\pm\kt{-+}\right)\nonumber\\
\kt{\Phi_{Bell\pm}} &=& \frac{1}{\sqrt{2}}\left(\kt{++}\pm\kt{--}\right).
\end{eqnarray}
These four kets provide an alternate complete basis to the two-party spin-1/2 system.  These can also be expressed in the direct sum basis
\begin{eqnarray}
\kt{\Phi_{EPR+}} &=& \kt{10}\nonumber\\
\kt{\Phi_{EPR-}} &=& \kt{00}\nonumber\\
\kt{\Phi_{Bell\pm}} &=& \frac{1}{\sqrt{2}}\left(\kt{11}\pm\kt{1-\!1}\right).
\end{eqnarray}
There are also other superpositions with maximum entanglement, such as
\begin{eqnarray}\label{case}
\kt{\Phi_{E\theta\pm}}  &=& \frac{e^{i\theta}}{\sqrt{2}}\left(\kt{+-}\pm i \kt{-+}\right)\nonumber\\
&=& \frac{e^{i\theta}}{\sqrt{2}}\left(e^{\pm i \pi/4}\kt{\Phi_{EPR+}}\pm e^{\mp i \pi/4} \kt{\Phi_{EPR-}}\right)\nonumber\\
\kt{\Phi_{B\theta\pm}}  &=& \frac{e^{i\theta}}{\sqrt{2}}\left(\kt{++}\pm i \kt{--}\right)\nonumber\\
&=& \frac{e^{i\theta}}{\sqrt{2}}\left(e^{\pm i \pi/4}\kt{\Phi_{Bell+}}\pm e^{\mp i \pi/4} \kt{\Phi_{Bell-}}\right).
\end{eqnarray}

To study dynamical entanglement in scattering systems, consider the time evolution operator $U(t_i,t_f)$ which evolves the two-party state from $t_i$ to $t_f$.  The initial and final systems are assumed to be non-interacting, or equivalently it is assumed that the interaction occurs during a finite time interval, so the S-matrix can be defined as $\hat{S} = \lim_{t\rightarrow \infty} U(-t,t)$.  If there were no symmetries, then the S-matrix could be any element of $U(4)$, the unitary $4\times 4$ matrices.  However, rotational invariance, expressed as $\hat{S} =U(u)\hat{S}U^\dag(u)$ for all $u\in\su$, puts strict constraints on the form of the S-matrix: 
\begin{equation}
\br{s'\chi'}S\kt{s\chi} = \delta_{s's}\delta_{\chi'\chi}e^{2i\delta_s},
\end{equation}
where $2\delta_s$ is the scattering phase shift and depends on the total spin.  This can be seen as a consequence of the Wigner-Eckart theorem applied to a scalar operator, but more fundamentally it arises because $[\hat{\mathbf{s}},\hat{S}]= 0$ and so they must be simultaneously diagonalizable. Therefore the direct sum basis states are eigenkets of the S-matrix.  The S-matrix applied to a general direct product state (\ref{prod}) can be written
\begin{equation}\label{sopspin}
\hat{S}\kt{\Phi_{12}} =\sum_{\chi'_1 \chi'_2} \sum_{s \chi} \bk{s \chi}{\chi'_1 \chi'_2}\bk{s \chi}{\chi_1 \chi_2}e^{2i\delta_s}\kt{\chi'_1 \chi'_2},
\end{equation}
where CGCs for the $\su$ have been used twice.

Choosing our coordinate systems as in (\ref{in-jm}), the general out-state is
\begin{equation}\label{out-jm}
\Phi' = S\Phi = e^{2i\delta_1}\cos\theta\kt{11} + \frac{1}{\sqrt{2}}e^{2i\delta_1}\sin\theta\kt{10} + \frac{1}{\sqrt{2}}e^{2i\delta_0}\sin\theta\kt{00}.
\end{equation}
This can be re-expressed in the direct product basis as
\begin{equation}\label{out-ss}
\Phi' = e^{2i\delta_1}\cos\theta\kt{++} + \frac{\sin\theta}{2}\left\{(e^{2i\delta_1} + e^{2i\delta_0})\kt{+-} + (e^{2i\delta_1} - e^{2i\delta_0})\kt{-+}\right\}.
\end{equation}

The entanglement of (\ref{out-ss}) can be calculated using (\ref{ent}) and the result is
\begin{equation}\label{entres}
E(\Phi') = 1 - \frac{1}{2}\log_2 \left( (1 + x)^{(1 + x)} (1-x)^{(1-x)}\right)
\end{equation}
where
\begin{equation}
x = \sqrt{ 1 - \sin^4\theta\sin^2(2\Delta\delta)},
\end{equation}
and $\Delta\delta= \delta_0 - \delta_1$.  The numbers $\lambda_\pm = \sqrt{(1 \pm x)/2}$ are the Schmidt coefficients for the state $\Phi'$.\cite{nielsen}  The entanglement is at a maximum $E=1$ when $x=0$ and it is at a minimum $E=0$ when $x=1$ and $\Phi'$ is separable.

To better understand the physical significance of this results, let us look at special cases in which no entanglement occurs:
\begin{itemize}
\item If $\theta = 0$, then the two incoming particles have aligned spins, the reaction takes place entirely in the $s=1$ channel, and there is no entanglement generated. The effect of the interaction is just a phase shift.
\item If $2\Delta\delta = 0$, then both channels ($s=0$ and $s=1$) acquire the same phase and no entanglement is generated.
\item If $2\Delta\delta = \pi$, then there is still no entanglement, although there is a spin-exchange in the term with $\chi_1 \neq \chi_2$.  For example, in the case of $\theta = \pi/2$, $\Phi = \kt{+-}$ becomes $\Phi' = \exp(2i\delta_1)\kt{-+}$.
\end{itemize}
To summarize, no entanglement occurs if only one UIR occurs in the decomposition of the in-state and even if multiple UIRs occur, the relative phase must be non-trivial. 

For all other in-states and phases, entanglement happens, but the system does not contain maximal entanglement $E=1$ except for the special case when $\theta = \pi/2$ and $2\Delta\delta = \pm \pi/2$.  Then the result of the interaction is a superposition of the singlet state $\kt{00}$ and triplet state $\kt{10}$:
\begin{eqnarray}
\Phi' &=& \frac{1}{2}e^{2i\delta_1} \left\{(1 \mp i)\kt{+-} + (1 \pm i)\kt{-+}\right\}\nonumber\\
&=& \frac{1}{\sqrt{2}}e^{2i\delta_1} \left(\kt{00} \mp i\kt{00}\right).
\end{eqnarray}
which is a special case of (\ref{case}). As shown by (\ref{entres}) every other possible state will have less entanglement than this, depending on the initial conditions and the difference in the phase shifts.

\section{Dynamical Entanglement in Scattering}

Now we consider the full scattering experiment: two non-realtivistic spin-1/2 particles are prepared separately by a preparation apparatus at time $t\rightarrow - \infty$.  The in-state $\phi^{in}$ is the direct product of the free-particle states, i.e. $\kt{\phi^{in}}=\kt{\phi^{in}_1}\otimes\kt{\phi^{in}_2}$.
The particles interact elastically and the S-matrix converts the in-state to the out-state $\kt{\phi^{out}} = \hat{S}\kt{\phi^{in}}$ with the same particles.  The observables, defined by the registration apparatus at time $t\rightarrow + \infty$, are best represented by an unentangled, density matrix $\rho^{out}= \rho^{out}_1 \otimes \rho^{out}_2$.  If the particles are identical fermions, the out-observable can be anti-symmetrized to account for spin-statistics. 

Since we consider non-relativistic elastic scattering, these vectors are elements of, and the operators act on, the UIR spaces of $\bar{G}$, the Galilean group $G$ extended by a operator corresponding to the central charge of mass.  The generators of the Lie algebra of the group $\bar{G}$ are $\{\hat{M}, \hat{H}, \hat{\mathbf{P}}, \hat{\mathbf{K}}, \hat{\mathbf{J}} \}$, respectively, the mass, the Hamiltonian, the momentum, the boost, and the angular momentum operators.  The UIRs of $\bar{G}$ are labeled three invariants $\{M, W, s\}$, respectively, the mass $M$, the internal energy $W$ (the difference between the total energy and the kinetic energy), and the intrinsic spin $s$.  The operators corresponding the the invariants (and other operators that appear later) can be given explicit expressions in term of generators, e.g. see Refs.~\cite{gal}.  The UIRs of $\bar{G}$ will be denoted $D^{\{M,W,s\}}=D^\mu$.

Basis kets are provided by the eigenkets of a maximum Abelian subalgebra, i.e. a complete set of commuting operators (CSCO).  Many CSCOs are possible; a physically useful choice for scattering is the three momentum operators $\hat{\mathbf{P}}$ and the spin component operator $\hat{s}_3$.  Eigenkets of these operators can be normalized like
\begin{equation}\label{mombas}
\bk{\vp' \chi'}{\vp \chi} = \delta^3{(\vp - \vp')}\delta_{\chi\chi'}.
\end{equation}
Single particle states are in correspondence with UIRs of $\bar{G}$.

Of all the transformations in the Galilean symmetry group, only rotations act on the spin degrees of freedom, i.e. for $u\in\su\subset G$:
\begin{equation}
U(u)\kt{\vp \chi} = U(u)\left(\kt{\vp}\otimes\kt{\chi}\right) = \kt{R(u)\vp}\otimes\sum_{\xi'}D^s_{\chi'\xi}(u)\kt{\chi},
\end{equation}
where $R(u)\in\mathrm{SO}(3)$ is the well-known two-to-one homomorphism and $D^s$ is the $2s+1$-dimensional UIR of $\su$.  The fact that the momentum degrees of freedom and spin degrees of freedom are separable will allow us to speak meaningfully about entanglement of the spins.  In contrast, quantifying spin entanglement in UIRs of the Poincar\'e is more difficult because the momentum and spin degrees of freedom are not separable~\cite{peres02,harsh_pra05}.

As usual, we can start from the direct product basis $\kt{\vp_1 \chi_1; \vp_2 \chi_2} = \kt{\vp_1 \chi_1}\otimes \kt{\vp_2 \chi_2}$, or we can reduce the direct product state into UIRs of the Galilean group using the Clebsch-Gordan series for the Galilean group.
Unlike the rotation group, the direct product of the Galilean group is not simply reducible, e.g. multiple copies of the UIR appear in the decomposition, which can be schematically written as:
\begin{equation}\label{galsum}
D^{\mu_1}\otimes D^{\mu_2} = \sum_W\sum_s d(\eta) D^{\{M_1 + M_2, W, s\} (\eta)},
\end{equation}
where $d(\eta)$ is a function that tells how many times the UIR $D^{\{M_1 + M_2, W, s\}}$ appears in the direct sum and some set of additional parameters $(\eta)$ label this degeneracy.

To make this more concrete, a convenient choice for basis kets of a UIR of $\bar{G}$ when $M_1 = M_2 = M_0$ is
\begin{equation}
\kt{\vp \chi (q l m s)} = \frac{q}{\sqrt{2}}\sum_{\chi_1 \chi_2}
 \int d^2\mathbf{\Omega}\, Y_{lm}(\mathbf{\Omega})\bk{s \chi}{\chi_1 \chi_2}\kt{\vp_1 \chi_1; \vp_2 \chi_2},
\end{equation}
where the momentums are related by $\vp_1 = (\vp + q\mathbf{\Omega})/2$ and  $\vp_2 = (\vp - q\mathbf{\Omega})/2$, $Y_{lm}(\mathbf{\Omega})$ are spherical harmonics as a function of relative momentum direction, and $\bk{s \chi}{\chi_1 \chi_2}$ are the $\su$ CGCs.  The characteristic invariants of a UIR that appears in the decomposition of the equal mass direct product are $\mu = \{M= 2M_0, W(q) = W_1 + W_2 + q^2/2M_0, s\}$.  The degeneracy labels $(\eta)$ for this case are the relative orbital angular momentum $l$ and its 3-component $m$ of the two particles.

Another useful basis choice  for studying scattering dynamics are the total angular momentum eigenket basis: that combines the orbital and spin angular momentum using CGCs of $\su$:
\begin{equation}\label{totangbas}
\kt{\vp j_3 (q l s j)} = \sum_{j = |l - s|}^{l + s} \sum_{j_3 = -j}^{+j}\bk{j j_3}{m \chi}\kt{\vp \chi (q l m s)}.
\end{equation}
This is still a UIR of $\bar{G}$ with invariants $\mu$, but is is a different orthogonalization of the $d(\eta)$-degenerate UIRs in (\ref{galsum}).  We will also use a basis in a \emph{reducible} subspace of the direct product representation:
\begin{equation}
\kt{\vp \chi_1 \chi_2 (q l m)} = \frac{q}{\sqrt{2}}
 \int d^2\mathbf{\Omega}\, Y_{lm}(\mathbf{\Omega})\kt{\vp_1 \chi_1; \vp_2 \chi_2}.
\end{equation}
In this basis, the momentum degrees of freedom have been combined but the spin degrees of freedom have not.

Because of the Galilean symmetry, the S-matrix commutes with the generators of the extended Galilean group.  Therefore, in the (\ref{totangbas}) basis, the S-matrix must be diagonal in $j$, $j_3$, $\vp$, $E(q) = W(q) + \vp^2/2M$ (or in the equal mass case $q$):
\begin{equation}
\br{\vp' j_3' (q' l' s' j')}S\kt{\vp j_3 (q l s j)} = S^j_{l's',ls}(q)\delta^3(\vp - \vp')\delta(q - q') \delta_{jj'}\delta_{j_3j_3'},
\end{equation}
where $S^j_{l's',ls}(q)$ is the reduced S-matrix.  It is a function of the deneracy parameters $(j,l,l')$, the UIR invariants $s,s'$, and the magnitude of the relative momentum $q$~\cite{gw}.

The amount of dynamical entanglement in the spin degrees of freedom in this most general case depends on the properties of $S^j_{l's',ls}(q)$.  As an example, we will treat the simplest case of central forces.  Then the S-matrix commutes with the orbital angular momentum and total spin angular momentum , $[\hat{\bf{L}}, \hat{S}] = [\hat{\bf{s}}, \hat{S}] = 0$, as well as the generators of $\bar{G}$.  So the S-matrix must be diagonal in $l$ and $s$ and the components $m$ and $\chi$:
\begin{equation}
\br{\vp' \chi' (q' l' m' s')}S\kt{\vp \chi (q l m s)} = S^{ls}(q)\delta^3(\vp - \vp')\delta(q - q') \delta_{ll'} \delta_{mm'} \delta_{ss'} \delta_{\chi\chi'},
\end{equation}
where $S^{ls}(q)$ is the fully-diagonalized, reduced S-matrix~\cite{gw}.  Conventionally, it is expressed in terms of the scattering phase shift as
\begin{equation}\label{redcens}
S^{ls}(q) = e^{2i\delta_{ls}(q)}.
\end{equation}

We now restrict ourselves to spin-unentangled, in-states in the $l$-th partial wave of the orbital angular momentum with well-defined orbital angular momentum component $m$ and energy $E(q)$:
\begin{equation}\label{scatbas}
\kt{\phi^{in}} = \sum_{\chi_1\chi_2} a_{\chi_1}b_{\chi_2} \kt{\vp \chi_1 \chi_2 (q l m)}.
\end{equation}
As previously mention, these  superpositions of the basis kets of more than one UIRs of $\bar{G}$ (i.e. they are not the basis kets of an irreducible representation space).  Applying the central force S-matrix (\ref{redcens}) to the in-state (\ref{scatbas}) gives 
\begin{equation}
\kt{\phi^{out}} =
\sum_{\chi_1\chi_2}\sum_{\chi'_1 \chi'_2} \sum_{s\chi} \bk{s\chi}{\chi'_1 \chi'_2}\bk{s\chi}{\chi_1 \chi_2}e^{2i\delta_{ls}(q)} a_{\chi_1}b_{\chi_2} \kt{\vp \chi'_1 \chi'_2 (q l m)}.
\end{equation}
For a given $\{\vp,q,l,m\}$, this has the exact form of (\ref{sopspin}).

What has been proved is that the dynamical entanglement of the spin degrees of freedom works in exactly the same was as if there were no other degrees of freedom (momentum) as long as (1) we consider a single partial wave of the orbital angular momentum with well-defined energy and momentum, (2) the particles have identical mass, (3) scattering interaction is generated by a central force, and (4) the particles are distinguishable (otherwise only totally antisymmetric channels occur).  This is possible because UIRs of the Galilean group are separable in the momentum and spin degrees of freedom.  However, unlike the spin ``scattering'' case, the phases will now also depend on the energy (or $q$) and the degeneracy parameter $l$, in addition to the total spin $s$.  Different assumptions about the symmetries of the interaction and the particles will give different results for maximum dynamical entanglement, but the concept has been shown.

\section{Conclusions}

What are the consequences of this calculation?  If an experiment is performed such that the particles enter detectors that select a partial wave of orbital angular momentum, then we can expect the amount of entanglement (measurable, perhaps, in some Bell-type experiment) to change as the energy of scattering is varied and as different partial waves are analyzed.  In any statistical system where two-body interactions are dominant, such results may be employable to look at the evolution of entanglement in the system.

More generally, these results show how the interplay of representation theory and interaction symmetries can limit the amount of entanglement that can be dynamically generated.  In the case of two spin-1/2 systems, maximum entanglement can only be generated by the dynamics if the scattering phase shifts are precisely tuned.

There are many further questions this work suggests.  How can this be extended to multi-party scattering?  Relativistic systems?  Are there general constraints on dynamical entanglement depending on the symmetry group or its particular representations?  What about mixed states? To answer these questions, Clebsch-Gordan techniques for the reduction of UIR products will have to be combined with results about more sophisticated entanglement measures.

\section*{Acknowledgments}

The author would like to thank Wilson Smith for inspiring this approach, Mark Bird for fruitful discussions, Amir Fariborz and the organizers of MRST 2005 for giving the author the opportunity to discuss this work, and American University for providing travel support.

\end{document}